\documentclass[12pt]{article}

\input{psfig}

\begin{document}

\title{Using Wavelets to reject background in Dark Matter experiments}

\date{}

\maketitle

\begin{center}
%List of authors
%C.E. Aalseth$^{b}$, F.T. Avignone
%III$^{b}$, R.L. Brodzinski$^{c}$, S. Cebri\'{a}n$^{a}$, E.
%Garc\'{\i}a$^{a}$, \\
%%D. Gonz\'{a}lez$^{a}$,
%%W.K. Hensley$^{c}$,
I.G. Irastorza$^{a}$\footnote{Corresponding author. E-mail:
Igor.Irastorza@cern.ch. Present address: CERN, EP Division,
CH-1211 Geneva 23, Switzerland}, A. Morales$^{a}$, S.
Cebri\'{a}n$^{a}$, E. Garc\'{\i}a$^{a}$, J. Morales$^{a}$, A. Ortiz de
Sol\'{o}rzano$^{a}$, S.B. Osetrov$^{b}$, J. Puimed\'{o}n$^{a}$, M.L.
Sarsa$^{a}$, J.A. Villar$^{a}$

%\footnote{Corresponding author:
%amorales@posta.unizar.es}, %, I.V. Kirpichnikov$^{d}$, A.A.
%Klimenko$^{e}$, H.S. Miley$^{c}$, \\ J. Morales$^{a}$, A. Ortiz de
%Sol\'{o}rzano$^{a}$, S.B. Osetrov$^{e}$, V.S. Pogosov$^{f}$, J.
%Puimed\'{o}n$^{a}$, J.H.\ Reeves$^{c}$, \\ M.L. Sarsa$^{a}$,
%%S. Scopel$^{a}$,
%A.A. Smolnikov$^{e}$, A.G. Tamanyan$^{f}$, A.A. Vasenko$^{e}$,
%S.I. Vasiliev$^{e}$, J.A. Villar$^{a}$
\end{center}

\begin{center}
\begin{em}

$^{a}$Laboratory of Nuclear and High Energy Physics, University of
Zaragoza, \\ 50009 Zaragoza, Spain
\\
%$^{b}$University of South Carolina, Columbia, South Carolina 29208
%USA
%\\
%$^{c}$Pacific Northwest National Laboratory, Richland, Washington
%99352 USA
%\\
%$^{d}$Institute for Theoretical and Experimental Physics, 117 259
%Moscow, Russia
%\\
$^{b}$Institute for Nuclear Research, Baksan Neutrino
Observatory,\\  361 609 Neutrino, Russia
%\\

%$^{f}$Yerevan Physical Institute, 375 036 Yerevan, Armenia \\

\end{em}
\end{center}


\abstract{A method based on wavelet techniques has been developed
and applied to background rejection in the data of the IGEX dark
matter experiment. The method is presented and described in some
detail to show how it efficiently rejects events coming from noise
and microphonism through a mathematical inspection of their
recorded pulse shape. The result of the application of the method
to the last data of IGEX is presented.

{\it PACs:} 95.35+d: 14.80.Mz

}

\section{Introduction}

Experiments searching for rare event phenomena, like those looking
for non-baryonic dark matter particles (axions or WIMPs),
supposedly filling a substantial part of the galactic haloes, or
those looking for the neutrinoless double beta decay have become
of most relevance in Particle Physics and Cosmology. To find the
nature of Dark Matter, the essential component of the Universe,
which in its non-baryonic form might be (according to recent
developments) responsible of $\sim$25-30 \% of the dark
energy/matter budget of a flat universe, is one of the big
challenges of modern Cosmology. On the other hand, the old subject
of nuclear double beta decay which has been the objective of many
experimental efforts as an unique tool to explore the nature of
the electron neutrino and to verify the lepton number
conservation, has become recently of even greater importance after
the confirmation that the neutrino has a non-zero mass. In fact,
the existence of neutrinoless double beta decay may provide
information on the absolute scale mass of the neutrino, its mass
pattern and possibly on CP violation in the leptonic sector if the
sensitivity of the experiments is able to reach the level of a few
meV for the Majorana neutrino effective mass upper bound.

The signals to be expected in the two selected examples of rare
event physics, have in common their low rates and so the main
challenge to achieve the required sensitivity is to disentangle
the signal from the various sources of background, which largely
hide the expected signals. Consequently, the first basic
requirement of these types of searches is to provide an ultra-low
background environment, as well as to design specific techniques
of shielding and background rejection.

In the particular case of dark matter searches, the expected
signal is concentrated in the lowest part of the spectrum
\cite{Mor99} often just above the threshold of the experiment.
When the effective threshold of these experiments is fixed by the
electronic baseline, random fluctuations of this baseline
--electronic noise-- can populate the energy range of interest
with noise events. Actually this electronic noise becomes a
primary source of background which is of special concern.

In this paper we describe a method which has been developed and
successfully applied to the last data set of the IGEX dark matter
experiment. In the section \ref{exp} we will describe briefly the
IGEX experiment, just to understand the needs that led to the
method presented in this paper. The details of the experiment can
be found in ref. \cite{Aal,Mor00} which contain also the
scientific results of IGEX. In sections 3 and 4 the wavelet method
itself and its calibration are described. We finish with the
conclusions in section 5.

\section{The Experiment}
\label{exp}

 The IGEX experiment \cite{Aal,Gon99}, optimized for detecting
$^{76}$Ge double-beta decay, has been described in detail
elsewhere. One of the IGEX detectors of 2.2 kg, enriched up to 86
\% in $^{76}$Ge, is being used to look for WIMPs interacting
coherently with the germanium nuclei.
% The COSME detector described
%below, is also operating in the same shield at Canfranc.
%The IGEX detectors were fabricated at Oxford
%Instruments, Inc., in Oak Ridge, Tennessee.
%Russian GeO$_2$ powder, isotopically enriched to 86\% $^{76}$Ge,
%was purified, reduced to metal, and zone
%refined to $\sim 10^{13}$ p-type donor impurities per cubic centimeter
%by Eagle Picher, Inc., in Quapaw, Oklahoma. The metal was then
%transported to Oxford Instruments by surface in order to minimize
%activation by cosmic ray neutrons, where it was further zone refined,
%grown into crystals, and fabricated into detectors.
%The COSME detector was fabricated at Princeton Gamma-Tech, Inc. in
%Princeton, New Jersey, using naturally abundant germanium. The
%refinement of newly-mined germanium ore to finished metal for this
%detector was expedited to minimize production of cosmogenic
%$^{68}$Ge.
%All of the cryostat parts were electroformed using a high purity OFHC
%copper/CuSO$_4$/H$_2$SO$_4$ plating system.
%The solution was continuously filtered to eliminate copper
%oxide, which causes porosity in the copper. A Ba(OH)$_2$ solution  was
%added to precipitate BaSO$_4$, which is also collected
%on the filter. Radium in the bath exchanges with the barium on the filter,
%thus minimizing radium contamination
%in the cryostat parts. The CuSO$_4$ crystals were
%purified of thorium by
%multiple recrystallization.
%The IGEX detector used for dark matter searches, designated RG-II, has
%a mass of $\sim2.2$ kg.
Its active mass is $\sim2.0$~kg, measured with a collimated source
of $^{152}$Eu.
%in the Canfranc Laboratory and is in agreement with
%the Oxford Instruments efficiency measurements.
The full-width at half-maximum (FWHM) energy resolution is
2.37~keV at the 1333~keV line of $^{60}$Co, and the low energy
long-term energy resolution (FWHM) is less than 1 keV at the 46.5
line of the $^{210}$Pb.

We refer to papers \cite{Mor00} and \cite{Morales:2001hj} where
the latest results of the experiment (regarding dark matter
searches) are presented and where the aspects related to the
set-up of the experiment, shieldings, the data acquisition system,
etc. are described. The threshold of the experiment is about 4 keV
and the raw background registered in the region just above
threshold is of a few tenths of counts per kg and day and per keV.
The background and threshold achievements provide the best
WIMP-nucleon cross-section versus WIMP mass ($\sigma$,m) exclusion
plot for spin-independent interactions ever obtained by an
experiment with no mechanism of nuclear/electron recoils
discrimination for background rejection. As stressed in
\cite{Mor00,Morales:2001hj}, the background level at low energies
is crucial to set the sensitivity of the experiment and this is
illustrated by the fact that a moderate reduction of the low
energy background would lead to a substantial improvement in the
final exclusion. To deal with the worrysome noise component in
this energy region trying its minimization, we developed the
analysis method described in the present paper. The acquisition
system for low energy data is implemented by splitting the
preamplifier output pulses and routing them through two Canberra
2020 amplifiers having different shaping times. These amplifier
outputs are converted using 200 MHz Wilkinson-type Canberra
analog-to-digital converters (ADC), controled by a PC through
parallel interfaces. The energy spectrum is built using the ADC
output of one of the lines. In addition, the output pulse of one
of the amplifiers is digitized and recorded for each event using a
800 MHz LeCroy 9362 digital scope. The digitized pulse encompasses
a time window of 500~$\mu$s with a sampling resolution of
1~$\mu$s. This recorded pulse shape is analyzed off-line with the
method described below and is formally referred as $f(t)$ in the
following.

\section{The method}

Wavelets are mathematical tools that have been proven to be
extremely useful in a number of different types of analysis
\cite{wavelets}. In particular, they have been applied to the
recognition and extraction of noise from certain sets of data.
From a mathematical point of view, the so-called wavelet
transforms can be viewed as extensions of the Fourier transforms,
but with the addition of sensitivity to local features of the
studied function. In fact, the set of functions which play the
role of the $e^{i\omega t}$ in the Fourier transform are now
$\psi_{a,b}(\frac{t-b}{a})$ derived from an original localized
null-area wavelet function $\psi(t)$ by means of translations and
contractions parameterized respectively by $b$ and $a$. Therefore,
for a fixed wavelet function $\psi(t)$, the biparametric wavelet
transform $[W_\psi f](a,b)$ of a given function $f(t)$ can be
defined as:

\begin{equation}\label{def_wavelet_trans}
  [W_\psi f](a,b) = \frac{1}{\sqrt{a}}\int_{-\infty}^\infty f(t)
  \psi\left(\frac{t-b}{a}\right)dt
\end{equation}

The above mentioned condition of null-area for $\psi(t)$, or
\emph{admisibility condition}, is derived from the requirement
that no information loss must occur in the transformation (i.e. it
must be invertible). Apart from this requirement, other properties
are desirable in certain contexts, like continuity, gaussian
shape, etc. Without entering into the mathematical details of the
wavelet theory \cite{wavelets}, and following a practical
approach, the "mexican hat" wavelet function $\psi(x)$ was chosen
because it fitted most suitably our requirements:

\begin{equation}\label{mexican_hat}
    \psi(x) \propto (1-x^2)\exp\left(-\frac12x^2\right)
\end{equation}

\begin{figure}[h]
\begin{center}
\hspace{0 cm} \psfig{figure=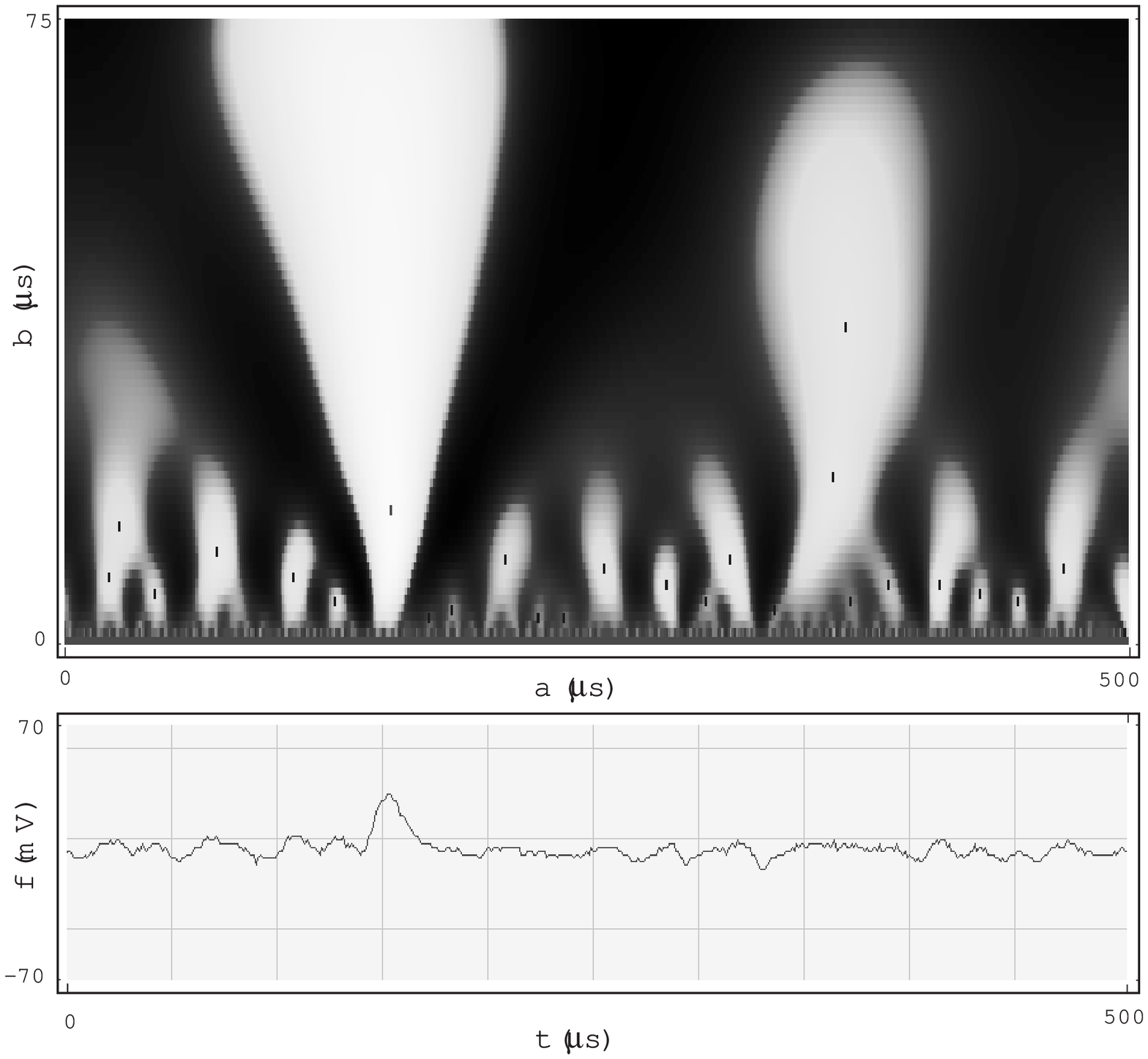,width=140mm}
\caption{In the lower part a sample pulse corresponding to an
event of about 4.5 keV is shown. In the upper part the
2-dimensional wavelet transform of the same pulse is presented.
Lighter regions correspond to highest values of the transform
function. The relative maxima are marked with dark spots. The
highest maximum is clearly related with the presence of the true
pulse, the others being caused by random fluctuations of the
baseline.\label{wavelet_sample}}
\end{center}
\end{figure}

As stated before, $f(t)$ is the pulse shape for each event as
recorded by the data acquisition system. Following expression
(\ref{def_wavelet_trans}), the wavelet transform function $[W_\psi
f](a,b)$ can be calculated numerically for each event. The
resulting function is continuous due to the properties of the
wavelet function used. So the relative maxima $(a_i,b_i)_i$ (where
$i$ goes from 1 to $N$, being N the total number of maxima) can be
calculated by means of an appropriate numerical algorithm. Let's
order the maxima $(a_i,b_i)$ according to the magnitude of the
wavelet transform in that maxima, $\omega_i \equiv [W_\psi
f](a_i,b_i)$, such that the first maximum $\omega_1$ is the
highest one. In some sense, these maxima contain the information
of the pulse. For instance, the position of the highest one in the
$(a,b)$ plane is determined by the width and position in the time
window of the event pulse itself, if present, and the others
$(a_i,b_i)_{i=2...N}$ are the result of random fluctuations of the
baseline along the time window encompassed by $f(t)$. That is
illustrated in figure \ref{wavelet_sample}, where in the lower
part a sample event pulse is shown (to give an idea of the
vertical scale, the pulse correspond to an event of about 4.5 keV,
i.e. just above our threshold). In the upper part of the figure
the 2-dimensional wavelet transform of the same pulse is
presented, lighter regions corresponding to higher values of the
$[W_\psi f](a,b)$. The relative maxima are also marked with darker
spots, the first one being clearly identified and corresponding to
the true pulse, and the others corresponding to random
fluctuations of the baseline.

\begin{figure}[h]
\begin{center}
\hspace{0 cm} \psfig{figure=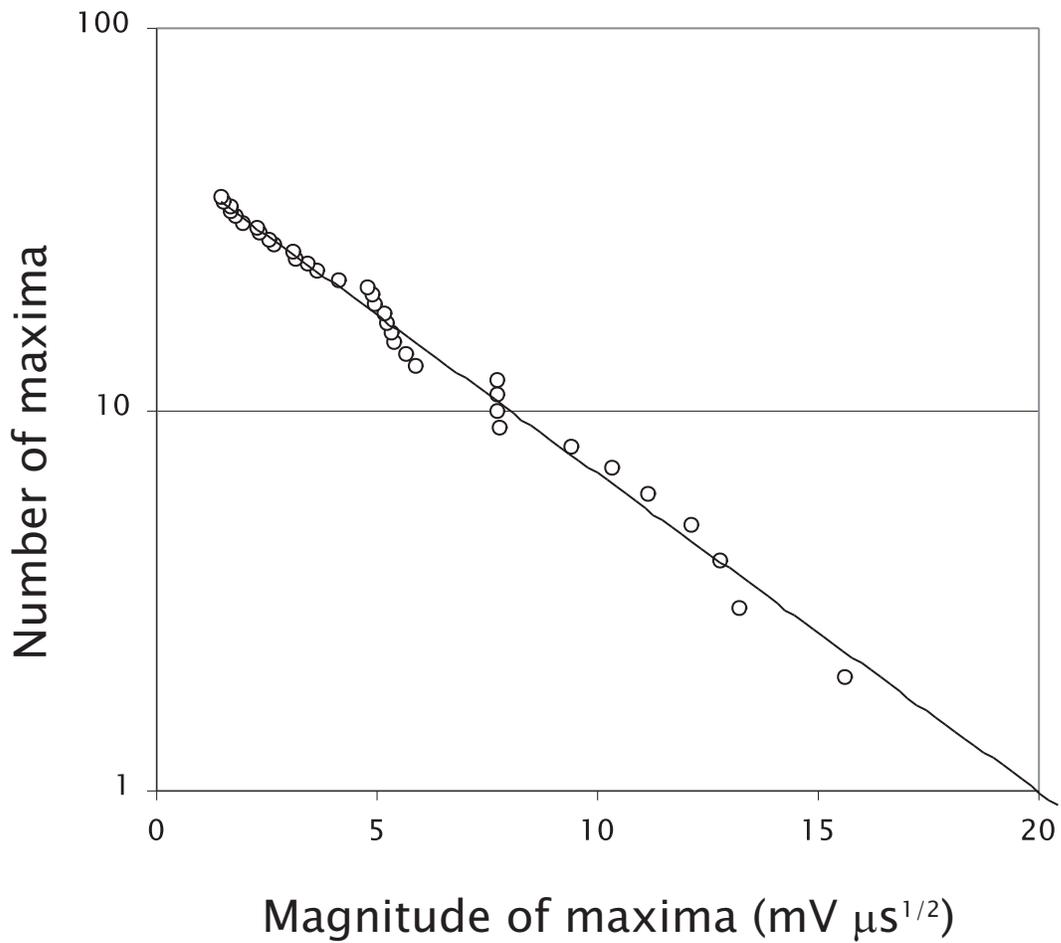,width=140mm}
\caption{Distribution of the maxima $\omega_i$ ($i$ going from 2
to $N$) of the wavelet transform for a sample event and their fit
to an exponential.\label{exponential}}
\end{center}
\end{figure}

The key idea of the method is based on the comparison between the
magnitude of the first maximum $\omega_1$ with the distribution of
the others $\{\omega_i \}_{i=2...N}$. In fact, the mentioned
distribution has been checked to be well modelled by a decaying
exponential (this fact being related to the kind of wavelet
function we are using). Figure \ref{exponential} shows such a
distribution for one event.

Now, the method consists of constructing the exponential
distribution $D(w)=ke^{-\alpha w}$ which follows the baseline
induced maxima in every event pulse shape and evaluate the
parameter $P_1$ of the first maximum, defined as the normalized
area of the distribution $D(w)$ which lies above the first maximum
$w_1$:

\begin{equation}\label{P1}
P_1=\frac{\int_{w_1}^{\infty} D(w)dw}{\int_{0}^{\infty}
D(w)dw}=e^{-\alpha w_1}
\end{equation}

This parameter $P_1$ is the final outcome of the method for each
event, and gives us, to a certain extent, an indication of the
probability of the main pulse to have been generated randomly by
the fluctuations of the baseline, being smaller for pulses farther
from a noise-like pulse. Therefore, if $P_1$ is large enough, the
event must be considered a noise fluctuation and should be
rejected, otherwise, the event is kept. The reference value to
decide if $P_1$ is large enough or not must be fixed by
calibration of the method, and is explained in the following
section.

\section{Calibration}

\begin{figure}[h]
\begin{center}
\hspace{0 cm} \psfig{figure=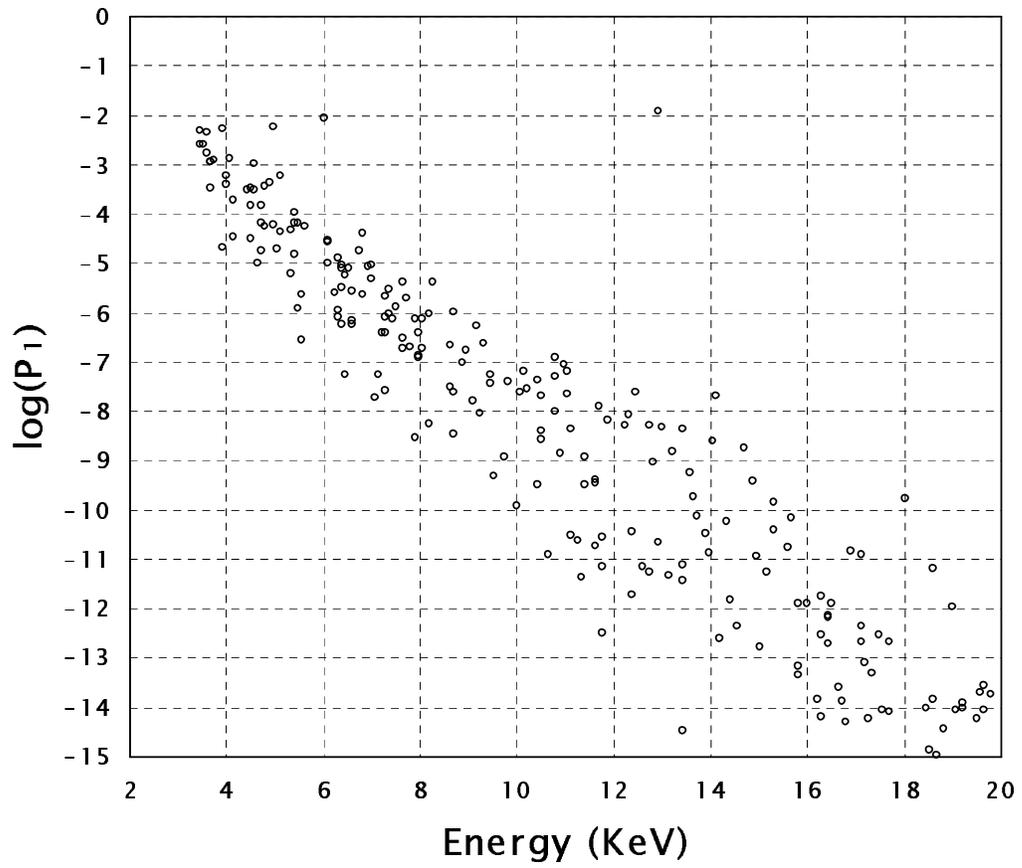,width=140mm}
\caption{Scatter plot showing the parameter $P_1$ assigned to each
 event by the wavelet technique (described in the text) versus
 energy for a calibration set of data. \label{wavelet_cal}}
\end{center}
\end{figure}

\begin{figure}[h]
\begin{center}
\hspace{0 cm} \psfig{figure=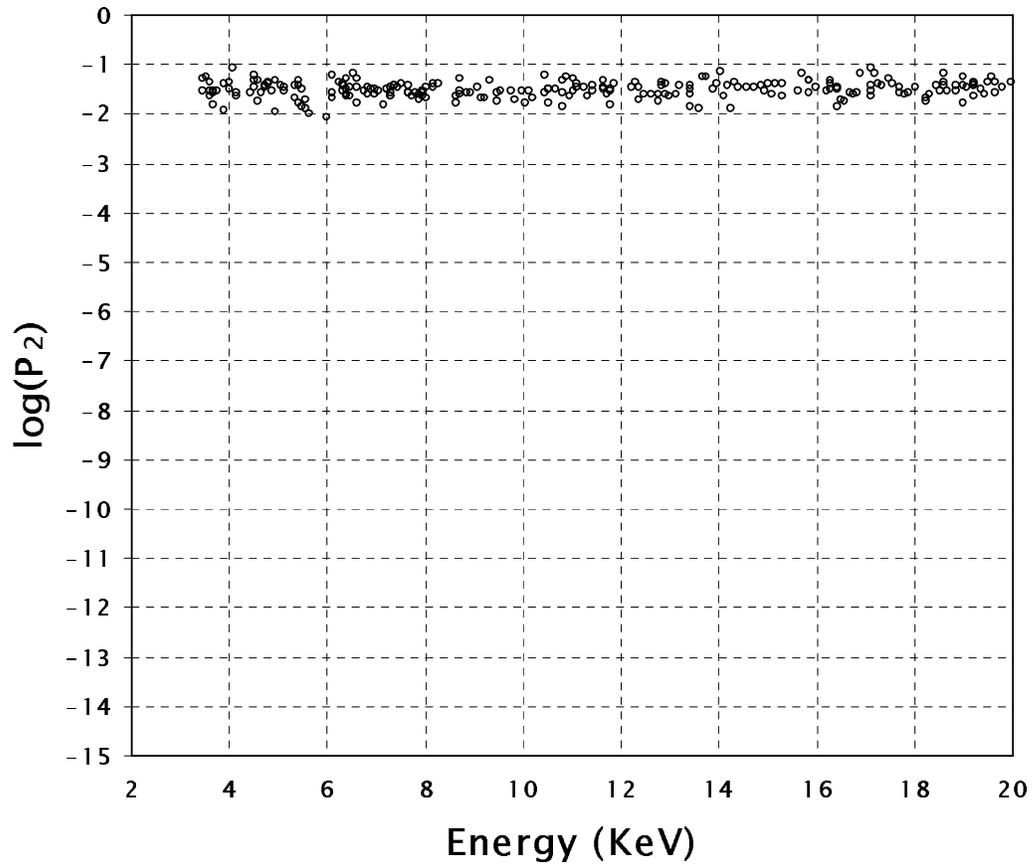,width=140mm}
\caption{Scatter plot showing the parameter $P_2$ related to the
second maximum of the wavelet transform versus the energy of the
event (for a calibration set of data).\label{wavelet_second}}
\end{center}
\end{figure}

\begin{figure}[h]
\begin{center}
\hspace{0 cm} \psfig{figure=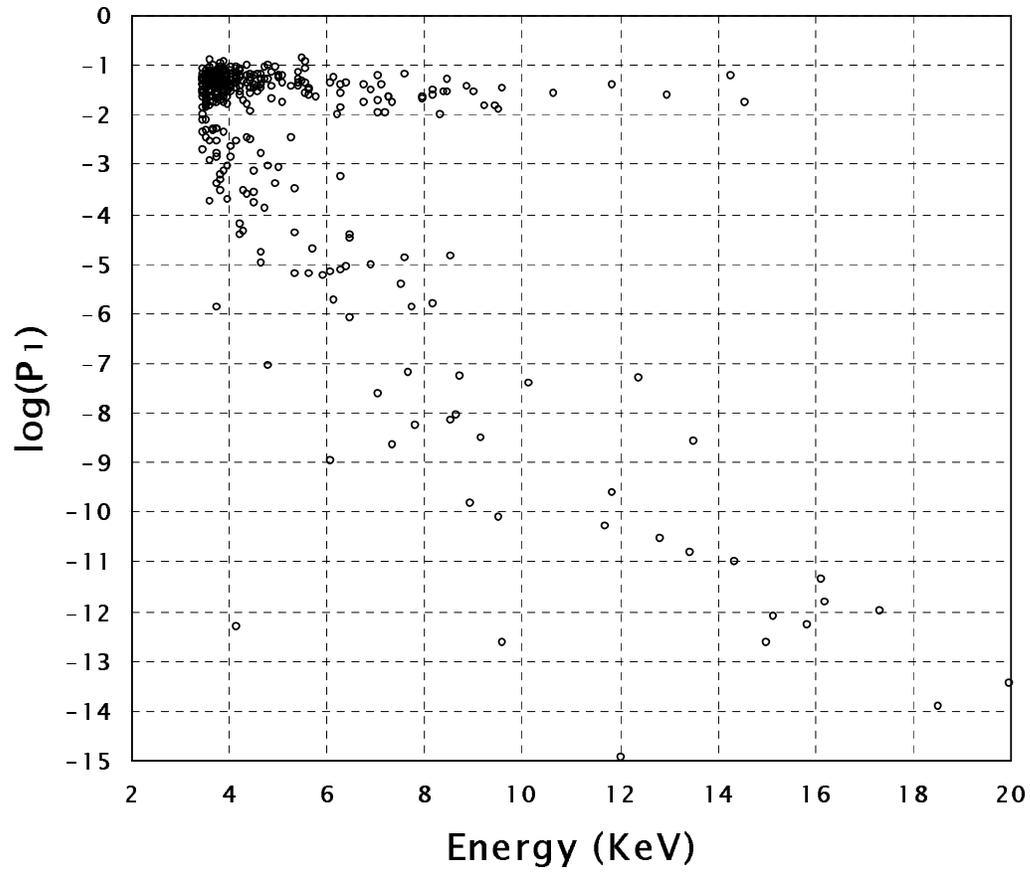,width=140mm}
\caption{Same as figure \protect\ref{wavelet_cal} but for
 background data. The populations of noise and data are well
 separated above 4 keV. \label{wavelet_bkg}}
\end{center}
\end{figure}

In order to fix the rejection criterion this method was applied to
a calibration set of data. The calibration was performed with a
$^{22}$Na source which was introduced inside the shielding through
a teflon probe producing a rate of events much higher than the
standard background. The source provides an almost pure sample of
"true" events, i.e. events not coming from noise. The result of
the method applied to this set of data is illustrated in figure
\ref{wavelet_cal} where the parameter $P_1$ (obtained as
previously explained) versus energy is presented for each event of
the calibration set. As expected, this parameter shows a clear
correlation with the event energy. It is interesting to see that
the same plot for the parameter $P_2$, defined similarly as $P_1$
but for the second maximum $w_2$, shown in figure
\ref{wavelet_second} lacks completely this correlation, as this
second maximum is related to some fluctuation of the baseline
other than the true pulse. The same type of plot obtained for a
background set of data is shown in figure \ref{wavelet_bkg}. In
this plot two different populations are clearly visible. One of
them follows the same energy dependence than that of the
calibration data and the other does not. This second type of
events are identified as noisy fluctuations of the baseline that
have triggered the acquisition system but no relevant pulse above
threshold is present. From these plots a criterion of $P_1<0.01$
can be defined to distinguish the two populations of noise and
data. Due to the limited statistics of the set of data available,
it is hard to quantify the energy dependence of the loss of
efficiency of the cut. However, from the presented plots one can
state that it is negligible for events which lay a little above
our threshold, whereas only in the very first keV above threshold
it could be only a few percent.

\begin{figure}[h]
\begin{center}
\hspace{0 cm} \psfig{figure=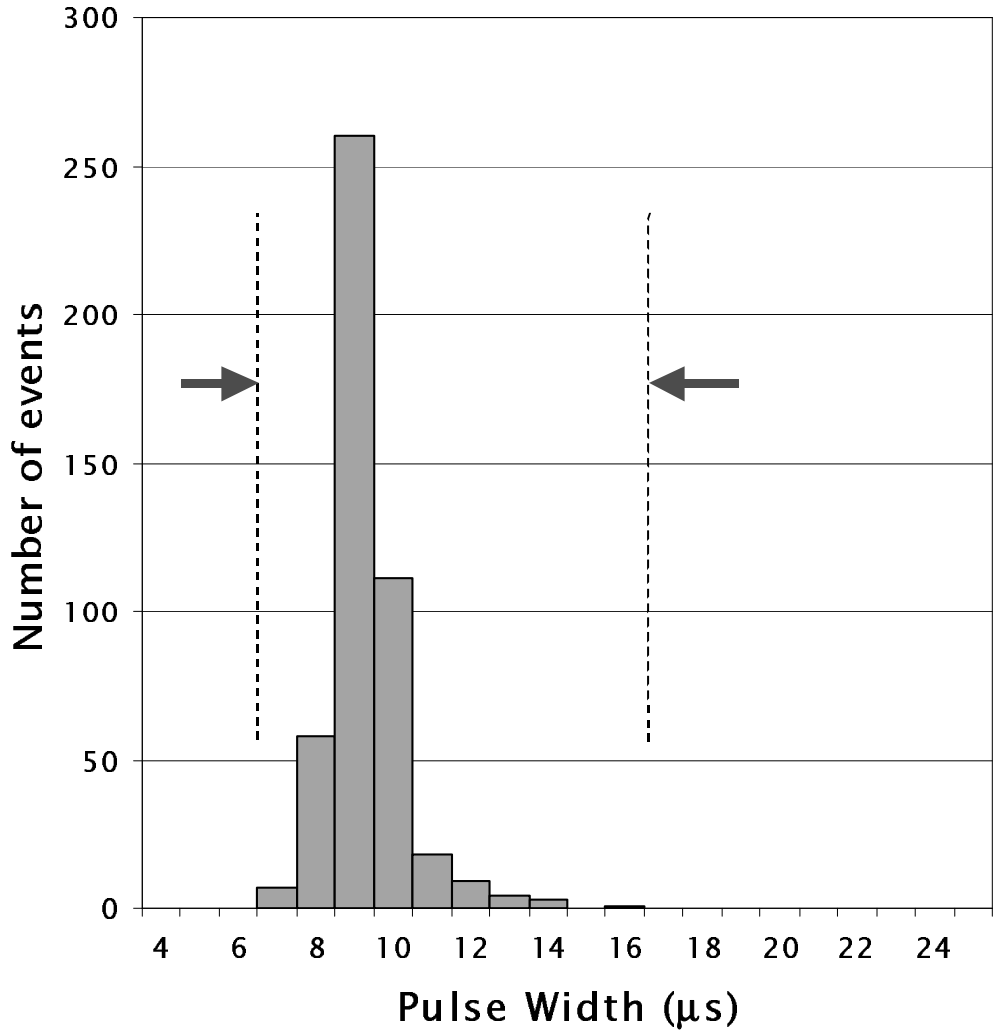,width=140mm}
\caption{Distribution of pulse widths for a calibration set of
data considering only events from 4 to 40 keV. All of them are in
the range 7 to 16 $\mu$s. \label{wavelets_width1}}
\end{center}
\end{figure}

\begin{figure}[h]
\begin{center}
\hspace{0 cm} \psfig{figure=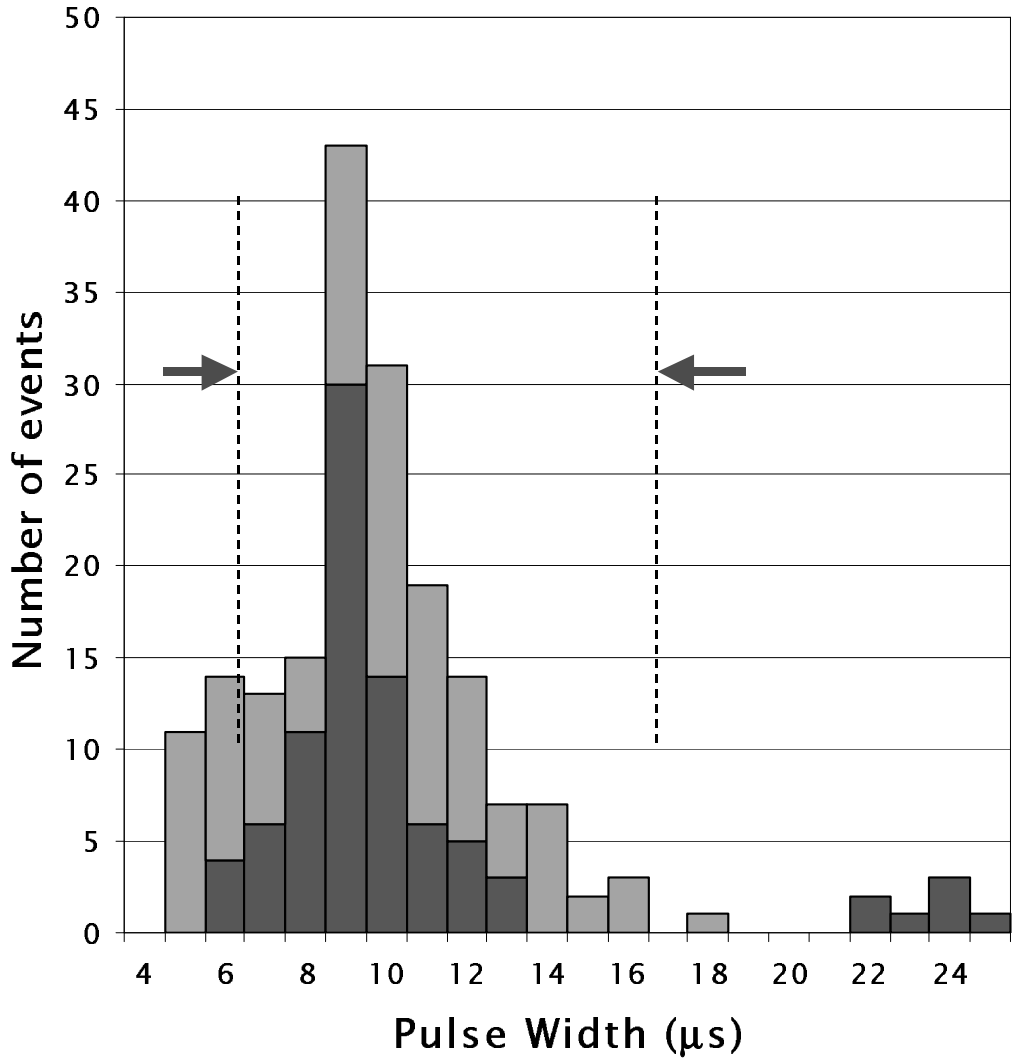,width=140mm}
\caption{Distribution of pulse widths for a background set of data
(considering only events from 4 to 40 keV), before and after the
rejection based on the criterion on the parameter $P_1$. It can be
appreciated some events with large width (above 16 $\mu$s) which
are not rejected by such criterion. \label{wavelets_width2}}
\end{center}
\end{figure}

Moreover, this method provides another way of rejection. The
calculation of the first maximum of the wavelet transform provides
the width of the corresponding pulse: $a_1$. This parameter is
tightly constrained for true pulses due to the shaping. As can be
seen in the figure \ref{wavelets_width1}, for our present set-up
all of the low energy pulse widths are between 7 and 16 $\mu$s.
This fact can be used to reject events with pulse widths outside
this range. This argument allows to reject a small population of
events which were not rejected by the previous method, fact that
is illustrated in figure \ref{wavelets_width2}. They correspond to
large lower-frequency fluctuations of the baseline presumably
connected with microphonism. Due to the fact that they are not
strictly random fluctuations in the sense previously considered,
they could not be rejected before. The loss of efficiency of this
cut is completely negligible.

\section{Conclusions}

\begin{figure}[h]
\begin{center}
\hspace{0 cm} \psfig{figure=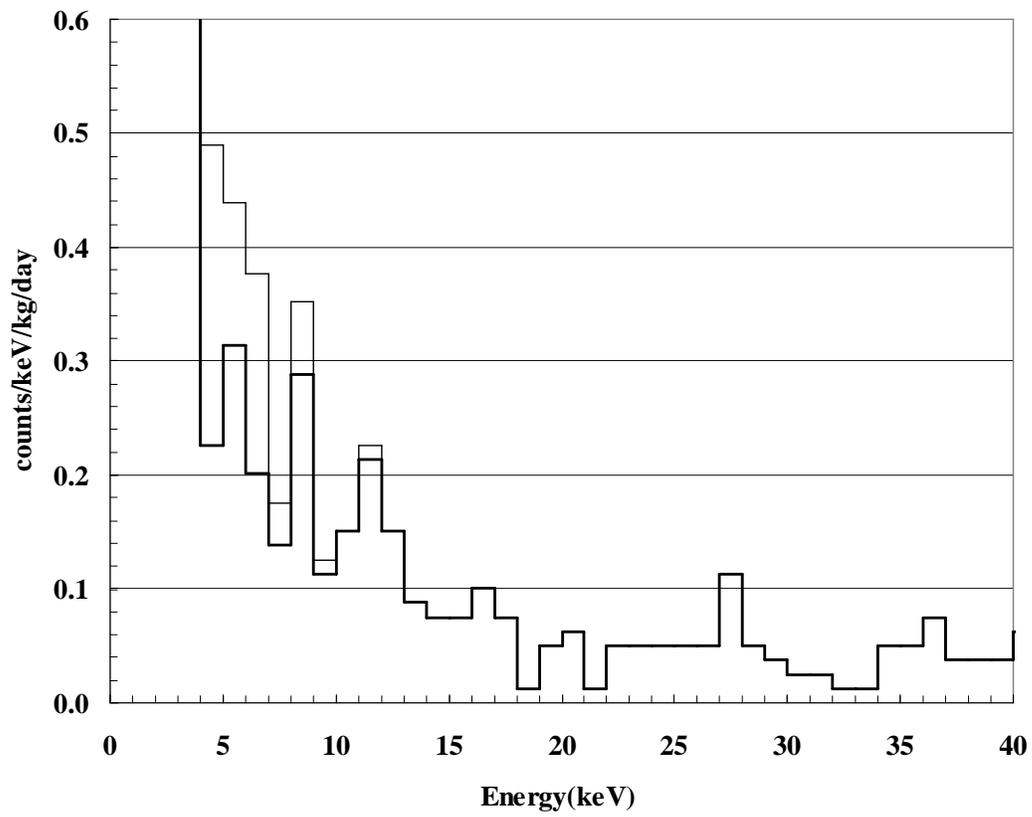,width=140mm} \caption{Low
energy spectrum of the last published 80 kg$\cdot$day of data of
IGEX before (thin line) and after (thick line) the application of
the wavelet rejection method. \label{spectrum}}
\end{center}
\end{figure}

We have presented a method based on wavelet techniques which is
able to efficiently reject electronic noise and microphonism from
the low energy --just above threshold-- region. The method
analyzes the recorded pulse shape of the event to extract
information about the pulse height in comparison to the
fluctuations of the baseline to set if the pulse can be assumed to
be randomly generated by the fluctuations. This method was
succesfully applied in the last data of the IGEX dark matter
direct search. In figure \ref{spectrum} we show the low energy
spectrum obtained with the last published 80 kg$\cdot$day of data
of IGEX \cite{Morales:2001hj} showing the effective rejection
achieved with the present method. About half of the background
between 4 and 7 keV and about 20\% between 7 and 10 keV was
identified as noise and eventually rejected.

\section{Acknowledgments}
This work has been partially supported by MCYT, Spain, under
contract No. FPA2001-2437. We are indebted to our IGEX colleagues
for their collaboration in other tasks of the experiment. I.G.I.
wish to thank Prof. J. Garay for his illuminating discussions on
the wavelet transform.

\newpage

\newpage

\newpage

\newpage

\newpage

\newpage

\newpage

\newpage

\end{document}